\documentclass[useAMS,usenatbib]{mn2e}

\usepackage{graphicx}

\def\rc{\mbox{$R_{\rm c}$}}
\def\rcl{\mbox{$R_{\rm cl}$}}
\def\mcl{\mbox{$M_{\rm cl}$}}

\def\ms{\mbox{$M_\odot$}}

\title[Hierarchical structures in the LMC and SMC]{Hierarchical structures in the 
Large and Small Magellanic Clouds}

\author[C. Bonatto and E. Bica]{C. Bonatto$^1$ and E. Bica$^1$\\
$^1$ Departamento de Astronomia, Universidade Federal do Rio Grande do Sul, Av. Bento 
Gon\c{c}alves 9500\\ 
Porto Alegre 91501-970, RS, Brazil}

\begin{document}

\pagerange{\pageref{firstpage}--\pageref{lastpage}}

\maketitle

\label{firstpage}

\begin{abstract}
We investigate the degree of spatial correlation among extended structures in 
the LMC and SMC. To this purpose we work with sub-samples characterised by different 
properties such as age and size, taken from the updated catalogue of Bica et al. or 
gathered in the present work. The structures are classified as star clusters or non-clusters 
(basically, nebular complexes and their stellar associations). The radius distribution 
functions follow power-laws ($dN/dR\propto R^{-\alpha}$) with slopes and maximum radius 
($R_{max}$) that depend on object class (and age). Non-clusters are characterised by 
$\alpha\approx1.9$ and $R_{max}\la472$\,pc, while young clusters (age $\la10$\,Myr)
have $\alpha\approx3.6$ and $R_{max}\la15$\,pc, and old ones (age $\ga600$\,Myr) have 
$\alpha\approx2.5$ and $R_{max}\la40$\,pc. Young clusters present a high degree of 
spatial self-correlation and, especially, correlate with star-forming structures, 
which does not occur with the old ones. This is consistent with the old clusters 
having been heavily mixed up, since their ages correspond to several LMC and SMC 
crossing times. On the other hand, with ages corresponding to fractions of the 
respective crossing times, the young clusters still trace most of their birthplace 
structural pattern. Also, small clusters ($R<10$\,pc), as well as small non-clusters 
($R<100$\,pc), are spatially self-correlated, while their large counterparts of both 
classes are not. The above results are consistent with a hierarchical star-formation 
scenario for the LMC and SMC. 
\end{abstract}


\section{Introduction}
\label{intro}

Star formation in the Milky Way and other galaxies is described as a (mass and size) 
scale-free, hierarchical process, in which turbulent gas forms large-scale structures 
with a mass distribution following a power-law. In essence, such a scale-free
process leads to a mass and size fractal distribution. As a consequence, young stellar 
groupings are clustered according to hierarchical patterns, with the great star complexes 
(associated with the $\sim10^7\,\ms$ superclouds) at the largest scales and the OB associations 
and subgroups, small loose groups, clusters and cluster subclumps (e.g. \citealt{Efremov95}) 
at the smallest. 
 
In several galaxies, the interstellar gas appears to follow a fractal structure ranging 
from the subpc ($\approx$ the current resolution limit) to the kpc scales; if star formation 
occurs preferentially at the densest regions, stars should form following such patterns 
(e.g. \citealt{ElmeElme01} and references therein). In this context, star clusters, 
formed at the core (i.e. the densest regions) of giant molecular clouds, can be taken 
as the unavoidable star formation product in a hierarchically structured gas (e.g. 
\citealt{Elme06}). A similar picture, in which star clusters are present in dense 
cores, emerges from numerical simulations that follow in time the collapse of gas clouds 
(e.g. \citealt{Walsh06}), which also occurs when effects of radiative feedback and magnetic 
fields are included (\citealt{Bate09}). 

\begin{figure*}
\begin{minipage}[b]{0.49\linewidth}
\includegraphics[width=\textwidth]{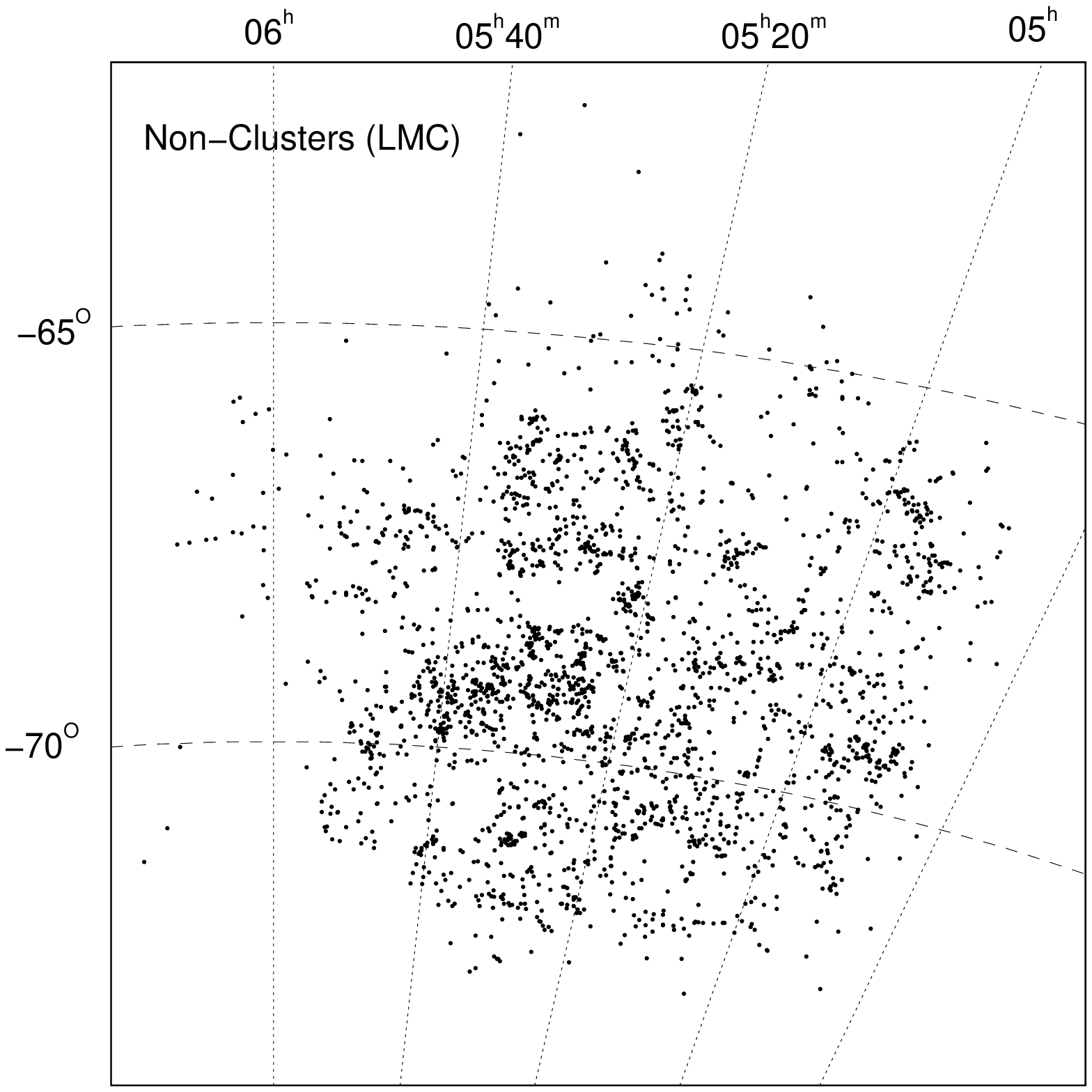}
\end{minipage}\hfill
\begin{minipage}[b]{0.49\linewidth}
\includegraphics[width=\textwidth]{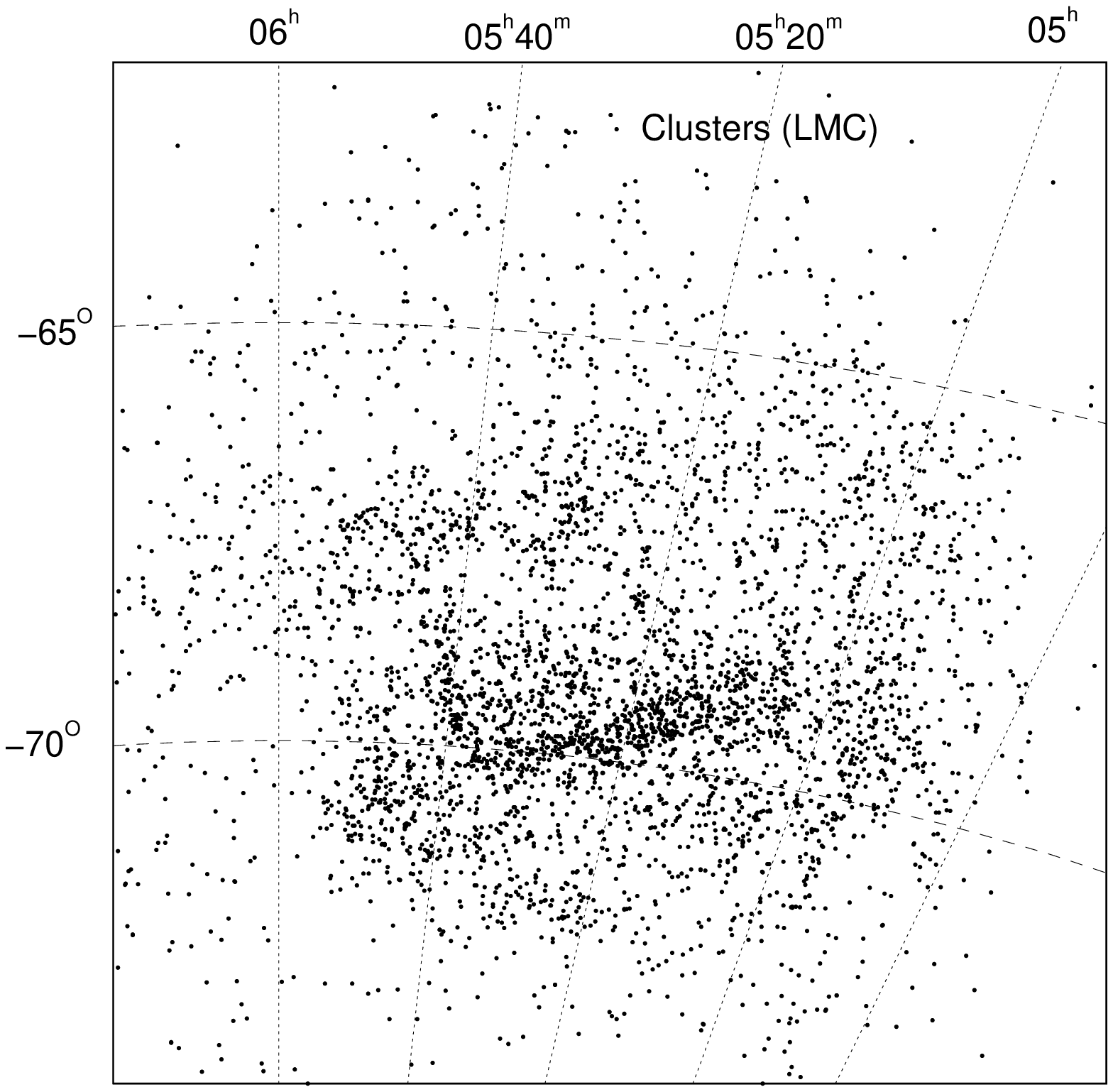}
\end{minipage}\hfill
\begin{minipage}[b]{0.50\linewidth}
\includegraphics[width=\textwidth]{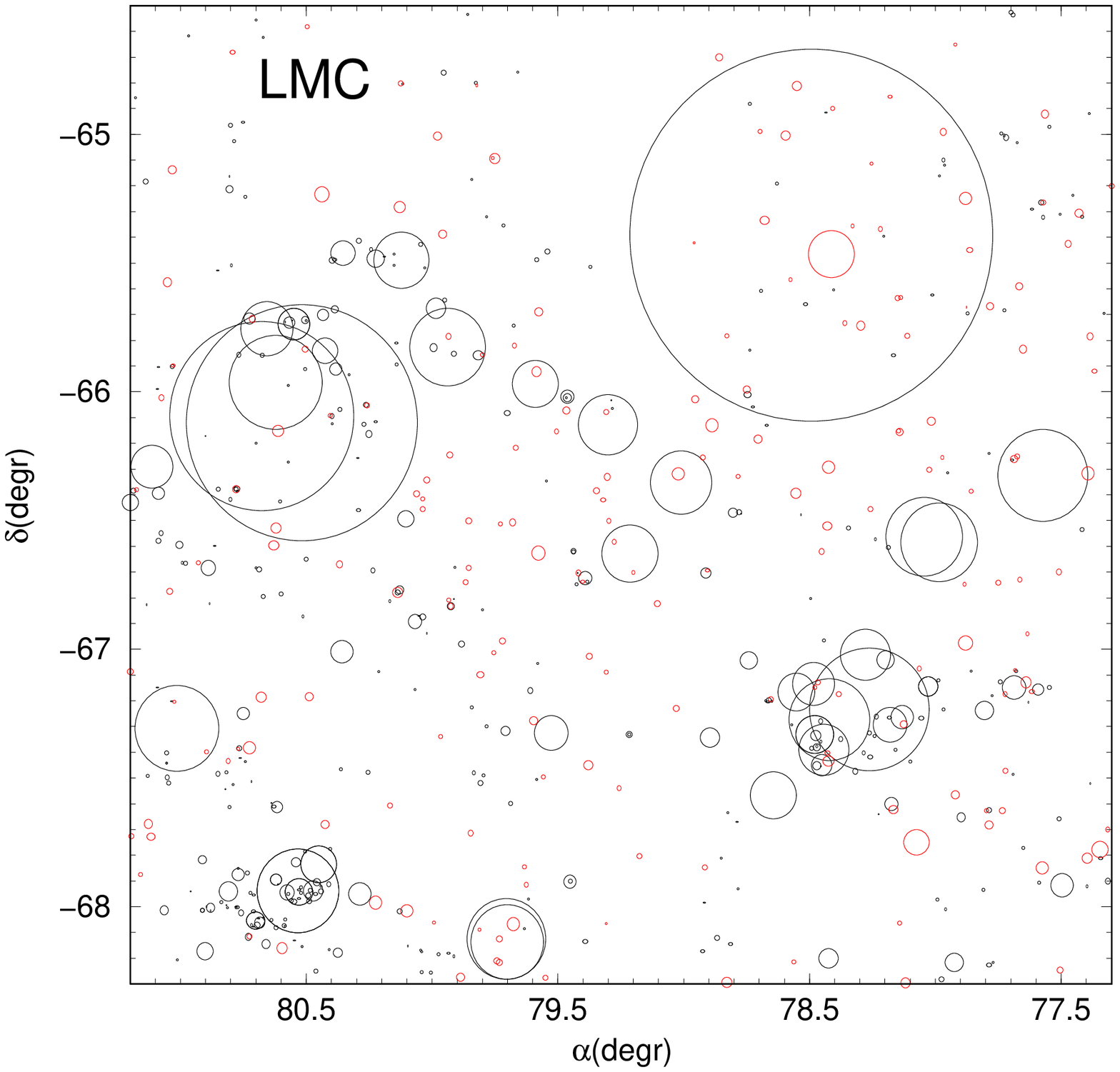}
\end{minipage}\hfill
\begin{minipage}[b]{0.50\linewidth}
\includegraphics[width=\textwidth]{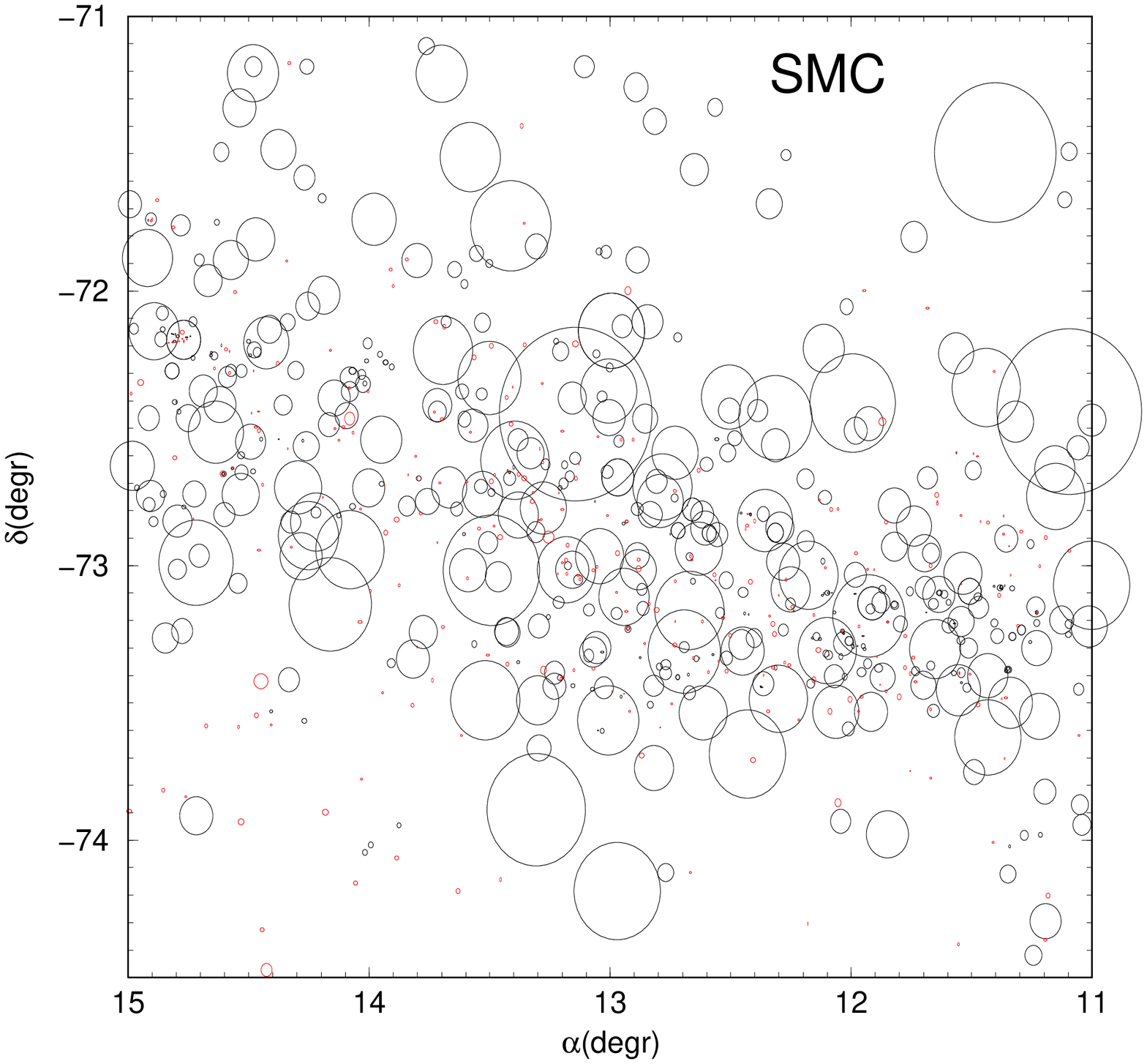}
\end{minipage}\hfill
\caption[]{Top: LMC non-cluster (Table~\ref{tab1}) structures are clumpier (left) than 
the clusters (right). Bottom: When cluster (red circles) and non-cluster (black) angular 
sizes are shown, hierarchical structuring appears to occur in these typical LMC (left) 
and SMC (right) fields. By far, most of the objects with a large angular size in the 
bottom panels are non-clusters}.
\label{fig1}
\end{figure*}

In a hierarchical scenario, the turbulent gas forms large-scale structures (clusters
and loose groups) with a mass distribution following a power-law of negative slope, 
i.e. $dN/dM\propto M^{-\beta}$, with $\beta\approx2$, consistent with the mass
distribution functions measured in several galaxies (\citealt{ElmeCon}).

Recent studies came up with robust evidence indicating that star-forming regions are 
indeed hierarchically structured, for instance in the nearby spiral galaxies M\,33 
(\citealt{Bastian07}), M\,51 (\citealt{Bastian05}), and NGC\,628 (\citealt{N628}), 
the Local Group dwarf irregular galaxy NGC\,6822 (\citealt{Karampelas09}), the 
Galactic disk (\citealt{GalDis}), and the Gould Belt (\citealt{Elias09}). 

Given the relative proximity, the Magellanic Clouds are an excellent environment to 
investigate the above issues. For instance, \citet{EfreElme98} found that the average 
age difference between pairs of LMC clusters increases as a function of their distance, 
which implies hierarchical star formation coupled to evolutionary effects. The angular 
correlation of LMC stellar populations for separations between 2\arcmin\ ($\sim30$\,pc) 
and 40\arcmin\ ($\sim550$\,pc) also implies large-scale hierarchical structure in current 
star formation (\citealt{HZ99}). The character of the LMC HI structure as a function of 
scale, the filamentary and patchy structures of the high- and low-emission regions respectively, 
suggest that most of the ISM is fractal, presumably the result of pervasive turbulence, 
self-gravity, and self-similar stirring \citealt{EKSS01}). More recently, \citet{Bast09} 
found a highly substructured and rapidly evolving distribution in the LMC stars. They suggest
that all of the original structure is erased in $\sim175$\,Myr (approximately the LMC crossing 
time), with small-scale structures mixing first. Similar conclusions apply to the SMC, in which 
stars appear to have formed with a high degree of (fractal) sub-structure, possibly imprinted 
by the turbulent nature of the parent gas; these structures are subsequently erased by random 
motions in the galactic potential on a time-scale of a crossing time through the galaxy 
(\citealt{GBE08}).

In this paper we investigate the degree of spatial correlation among the different kinds 
of LMC and SMC extended structures listed in the updated catalogue of \citet{UpCat}, 
together with its relation to star formation. We also study properties of their size 
distribution functions. Only two wide-apart age ranges are used for spatial correlation 
purposes: {\em (i)} very young objects (not older than $\sim20$\,Myr, and probably younger 
than $\sim10$\,Myr), which encompass clusters related to nebular emission and associations 
related or not to emission, as classified and catalogued from sky survey plates by \citet{UpCat} 
and references therein, and {\em (ii)} old clusters (older than $\sim600$\,Myr). According to 
our definition, the dynamical age of the very young clusters is lower (Sect.~\ref{2pcf}) than 
the crossing time (of the host galaxy), while for the old ones it corresponds to several 
crossing times, which is important for interpreting the spatial correlation in different
time periods. Clusters within the wide age range $\approx20-600$\,Myr are not used in the 
spatial correlation analysis (Sect.~\ref{2pcf}). 
   
Only the Magellanic Clouds have so far such a deep, homogeneous information on star 
clusters, associations and nebulae. Exceptions are some neighbouring dwarf galaxies 
that have been surveyed and are {\em (i)} featureless (Ursa Minor), {\em (ii)} contain 
a few globular clusters (Fornax) or, {\em (iii)} star-forming events like in the Clouds 
(e.g. NGC\,6822 - \citealt{Karampelas09}).

This paper is organised as follows. In Sect.~\ref{UpCat} we briefly discuss the updated
Magellanic System catalogue. In Sect.~\ref{OSC} we describe the selection criteria for
star clusters older than the Hyades. In Sect.~\ref{DistribFunc} we discuss the size (and 
mass, for star clusters) distribution functions of the different classes of objects. In 
Sect.~\ref{2pcf} we examine the spatial correlation of the different structures by means 
of two-point correlation functions. Concluding remarks are given in Sect.~\ref{Conclu}.

\section{The updated MC Catalogue}
\label{UpCat}

Properties of the updated MC catalogue are fully discussed in \citet{UpCat}. We recall
here the basic statistical properties. Taking the LMC, SMC and the Bridge together, the 
updated catalogue contains, respectively, 3740 classical star clusters, 3326 associations, 
1445 emission nebulae, and 794 HI shells and supershells. With the recent additions and 
cross-identifications, \citet{UpCat} contains about 12\% more objects than those 
in \citet{BSDO99} and \citet{BD00} together.

\begin{table*}
\caption[]{LMC and LMC Extended Object Properties}
\label{tab1}
\renewcommand{\tabcolsep}{1.0mm}
\renewcommand{\arraystretch}{1.25}
\begin{tabular}{lcrcccrcccrccl}
\hline\hline
       &         &\multicolumn{3}{c}{LMC}&&\multicolumn{3}{c}{SMC}&&\multicolumn{3}{c}{LMC$+$SMC}\\
       \cline{3-5}\cline{7-9}\cline{11-13}\\
Class  & Age & N & $R_{max}$&$\alpha$&& N & $R_{max}$&$\alpha$&& N & $R_{max}$&$\alpha$&~~~~~~~Comments\\
       &(Myr)&   & (pc)     &        &&   &   (pc)   &        &&   &   (pc)\\
\hline
C&Any     & 2268 & 38 & $3.53\pm0.22$ && 456 & 30 & $3.01\pm0.24$ && 2724 & 38 & $3.60\pm0.23$ & Ordinary cluster\\
CN&$\la10$&   81 & 25 & $3.85\pm0.62$ &&   9 &  8 & $2.76\pm0.58$ &&   90 & 25 & $3.73\pm0.67$ & Cluster in nebula\\
CA&$5-20$ &  738 & 21 & $3.19\pm0.59$ && 110 & 11 & $2.59\pm0.89$ &&  848 & 21 & $3.10\pm0.57$ & Cluster similar to assoc.\\
AC&$10-30$& 1185 & 32 & $4.42\pm0.24$ &&  60 & 14 & $2.44\pm0.17$ && 1245 & 32 & $4.11\pm0.21$ & Assoc. similar to cluster\\
NC&$\la5$ &  183 & 16 & $3.52\pm0.30$ &&  72 &  8 & $4.15\pm0.66$ &&  255 & 16 & $3.62\pm0.26$ & Nebula w/prob. emb. cluster\\
\hline
Clusters& & 4455 & 38 & $3.29\pm0.22$ && 707 & 30 & $3.04\pm0.23$ && 5162 & 38 & $3.20\pm0.19$ & C+CN+CA+NC+AC\\
\hline
A   & $\la30$& 1476 & 171 & $2.10\pm0.15$ && 130 & 292 & $2.23\pm0.23$ && 1606 & 292 & $2.21\pm0.14$ & Ordinary association\\
AN  & $\la10$&  217 & 262 & $1.80\pm0.11$ &&  39 &  62 & $2.07\pm0.30$ &&  256 & 265 & $1.70\pm0.11$ & Association w/nebular traces\\
NA  & $\la5$ &  817 & 472 & $1.75\pm0.06$ && 169 & 283 & $2.08\pm0.17$ &&  986 & 472 & $1.73\pm0.07$ & Nebula w/embedded assoc.\\
DAN$^\dagger+$DNC$^\dagger$& $\la5$ &   77 & 400 & $1.15\pm0.09$ &&  33 & 144 & $1.02\pm0.16$ &&  110 & 400 & $1.05\pm0.09$ & Decoupled structures\\
\hline
Non-clusters&& 2587 & 472 & $1.94\pm0.06$ && 371 & 288 & $2.04\pm0.13$ && 2958 & 472 & $1.89\pm0.06$ & A+AN+NA+DAN+DCN\\
\hline
SNR         &---&   52 &  78 & $0.86\pm0.10$ &&  22 &  38 & $0.50\pm0.57$ &&   74 &  78 & $0.85\pm0.13$ & Supernova remnants\\
HI shells   &---&  124 & 472 & $3.43\pm0.40$ && 545 & 482 & $2.88\pm0.20$ &&  794 & 477 & $2.82\pm0.05$ & HI shells and supershells\\
\hline
\end{tabular}
\begin{list}{Table Notes.}
\item N is the number of objects; $\alpha$ is the power-law slope ($\phi(R)=dN/d\,R\propto 
R^{-\alpha}$) fitted to the large radii range (Sect.~\ref{DistribFunc}); $R_{max}$ is the 
maximum radius measured in each class. ($\dagger$) - small cluster or association in large 
nebula. DCN and DAN: the nebular and stellar component of the objects can be distinguished
on Sky Survey plates.
\end{list}
\end{table*}

Especially in view of the spatial correlation analysis (Sect.~\ref{2pcf}), in this 
paper we restrict the object selection to the LMC and SMC, not including Bridge or 
extended Wing structures. A census of the LMC and SMC extended structures is provided 
in Table~\ref{tab1}, separated according to object class and including the probable 
age range. We note that, based on similarities observed in the size distributions 
(Sect.~\ref{DistribFunc}), in the present paper we include the AC and NC classes 
(relatively young objects) into the cluster classification, thus resulting in a 
higher number of such objects than quoted in \citet{UpCat}. Besides the latter two 
classes, the cluster classification also contains the C, CN and CA classes. As non-clusters 
(structures mostly associated with star formation environments) we take the A, AN, NA, 
DAN and DNC classes (see Table~\ref{tab1} notes). The SNR and HI shells are not used
because they are object classes apart and their size distribution functions are 
significantly different from those of the clusters and non-clusters (Fig.~\ref{fig4}). 

Figure~\ref{fig1} (top panels) shows the angular distribution of the 4455 LMC clusters 
(left) and 2587 non-clusters (right). Both kinds of structures trace well-known LMC 
(and SMC) structures (e.g. \citealt{UpCat} and references therein). It is also clear 
that the non-clusters appear to present a high degree of spatial correlation, with 
most of them tightly clumped together. This applies as well to the clusters, but to 
a lesser degree, because young and old clusters present significantly different
levels of spatial self-correlation, the latter being essentially non-correlated 
(Sect.~\ref{2pcf}). 

When the angular sizes are considered (bottom panels), we see that most structures
are arranged according to complex patterns, with sub-structures located inside 
larger ones.

\section{Old star clusters}
\label{OSC}

The identification, characterisation and spatial distribution of old star clusters
in the Clouds has been a major concern throughout decades (e.g. \citealt{Hodge1960};
\citealt{Hodge1982}; \citealt{Bruck75}; \citealt{vdBergh1981}; \citealt{Bica96}).
By old or red star clusters we mean those older than the Hyades\footnote{On (blue)
sky surveys, both a Hyades-age and a much older cluster would appear to consist of 
a considerable number of stars of about the same magnitude. Visually, they would
look pretty much the same. Young clusters, in contrast, are dominated by just a 
few very bright stars, essentially those at the top of the main-sequence turnoff.} 
($\approx630$\,Myr), or intermediate age clusters (IACs) up to classical globular 
cluster ages. We adopted the definition of old star cluster by \citet{JP94} and 
\citet{Friel95}. Magellanic Cloud clusters about this age appear to show dynamically 
evolved surface density profiles (e.g. \citealt{MG03a}; \citealt{MG04}; \citealt{Carva08}). 
Since the cluster age distribution function drops significantly with age (see, e.g., 
Fig.~2 of \citealt{deGG09} for the Magellanic Clouds), the presently adopted old cluster 
definition encompasses a statistically more significant sub-sample (Sect.~\ref{Cnstru}) 
than what would result for, e.g. clusters older than 1\,Gyr.

Clusters are expected to mix up by random motions under the galactic potential 
on a time-scale of a crossing time that, for the SMC, is of the order of 75\,Myr 
(\citealt{GBE08}), and about twice that value for the LMC\footnote{As a caveat we 
note that these dynamical timescales are essentially based on random cluster orbits. 
However, there is kinematical evidence suggesting that the LMC cluster system rotates 
as a flattened disk, but the disk geometry and systemic velocity appear to be different 
for young and old clusters (e.g. \citealt{FIO83}; \citealt{SSOH92}; \citealt{Groch06}). 
Indeed, some studies of the intermediate age and old populations have found that the 
velocity dispersion increases with age (e.g. \citealt{HWR91}; \citealt{SSOH92}; 
\citealt{Graff00}). In any case, the dynamical timescales may be longer than those 
used in the present paper.}. Thus, the old clusters as defined above have ages that 
correspond to several crossing times of the respective galaxy, and any memory of the 
clumpy structures where they were born should have been erased. In this sense, they 
can be used as probes of the long-term behaviour of the cluster spatial correlation 
(Sect.~\ref{2pcf}). 

\subsection{Short cluster history in the Clouds: towards taking the census of the total 
population?}
\label{ShortCH}

\citet{Kron56} and \citealt{Lindsay58} discovered luminous and intermediate luminosity 
clusters in the SMC using plate material. \citet{HodgeWri1974} and \citet{Bruck75} 
discovered intermediate and low luminosity clusters, while \citet{Hodge86} discovered 
even fainter ones by means of 4m telescope plates. \citet{BH95} discovered low 
luminosity clusters on sky survey plates, while \citet{Pietr98} discovered low 
luminosity clusters by means of CCD imaging.

\citet{Hodge1960} identified 35 luminous old clusters by means of non-calibrated CMDs, 
of which 11 were discoveries. \citet{ShaLind63} and \citet{LW63} discovered 
most of the luminous and intermediate-luminosity clusters in the LMC, the latter work 
being dedicated to the outer parts. \citet{HodSex66} discovered intermediate-luminosity 
clusters, while \citet{Hodge1988} low-luminosity ones with 4m-telescope plates. 
\citet{OHSC1988} discovered low-luminosity clusters in the outer parts. \citet{Kontizas90} 
discovered additional intermediate and low-luminosity clusters, while \citet{BSDO99} 
discovered a large number of low luminosity clusters on sky survey plates. \citet{Pietr99}  
discovered low-luminosity clusters in the LMC  with CCD observations.

\citet{UpCat} and references therein have cross-identified these catalogues and a number of 
other studies, and is particularly suitable  as a starting point for a deeper new survey 
such as the Visible and Infrared Survey Telescope for Astronomy (VISTA)\footnote
{http://www.eso.org/gen-fac/pubs/messenger/archive/no.127-mar07/arnaboldi.pdf}. It also 
provides a mean to properly acknowledge previous discoveries and to unambiguously establish
new cluster findings.

\citet{Santiago98} serendipitously detected two faint clusters in an LMC bar field using HST. 
The clusters have masses comparable to those of Galactic open clusters and ages in the range 
$200-500$\,Myr. The clusters are extremely faint on DSS and XDSS images, which suggests that 
the Clouds might harbour an important open cluster counterpart population. Besides being a
powerful tool to explore probable red brighter and intermediate-luminosity star clusters in 
the Clouds (Table~\ref{tab2}), VISTA will be essential also to detect such a possible population 
of open cluster counterparts, and estimate their age distribution.

Based on the broad-band UBVR photometry of \citet{Hunter03}, \citet{deGA06} derived 
absolute values of age and mass for a sample of LMC star clusters to study the cluster 
formation rate, their characteristic disruption time-scale and the cluster mass
function in different mass ranges. The same sample was used for further investigation 
of the cluster formation rate and the disruption time-scale by \citet{PdG08}. The same 
method was applied to a sample of SMC clusters by \citet{deGG08} to study the {\em
infant mortality}. Our approach in this paper differs in several ways, since we intend
to build statistically significant samples of clusters (as well as associations and 
emission nebulae) characterised by very different age ranges.

\subsection{Construction of the present sample of old clusters}
\label{Cnstru}

We compiled ages from the literature later than 1988, as determined from CMDs. There are 85 
and 202 old clusters with ages derived from CMDs for the SMC and LMC respectively, and they 
are provided in Table~\ref{tab2}\footnote{Given the large number of star clusters (667),
Table~\ref{tab2} is available only in electronic format. Here we provide an excerpt, for
illustrative purposes.}. Columns~1 to 8 of this table contain the same information as the 
general catalogue (\citealt{UpCat}). We now introduced additional columns that provide the
age determination method (Col.~9), $\log(Age)$ (Col.~10), and the relevant references for
the age (Col.~11). Note that several references are compilations themselves, so more references 
are therein.  

\begin{table*}
\caption[]{Old SMC and LMC clusters inferred from different methods - Excerpt}
\label{tab2}
\renewcommand{\tabcolsep}{1.33mm}
\renewcommand{\arraystretch}{1.25}
\begin{tabular}{lrrcccccccl}
\hline\hline
Designations&$\alpha[J2000]$&$\delta[J2000]$&Class&a&b&PA&Classification& Method&log(Age)&Ref\\
           &($hms$)&(\degr\arcmin\arcsec)&&(\arcmin)&(\arcmin)&(\degr)&&&(yr)\\      
(1)        & (2)   &   (3) & (4) & (5) & (6) & (7) & (8) & (9) & (10) & (11) \\
\hline
\multicolumn{11}{c}{SMC Star Clusters}\\
\hline
AM-3,ESO28SC4  &                23:48:59 & -72:56:43 & C & 0.90 & 0.90 & - & Old IAC          & CMD & 9.74 & R3\\
L1,ESO28SC8    &                 0:03:54 & -73:28:19 & C & 4.60 & 4.60 & - & Globular Cluster & CMD & 9.95 & R12\\
   "           &                         &           &   &      &      &   &                  & CMD & 9.88 & R26\\
L2             &                 0:12:55 & -73:29:15 & C & 1.20 & 1.20 & - &                  & PLA & Red  & R55\\
L3,ESO28SC13   &                 0:18:25 & -74:19:07 & C & 1.00 & 1.00 & - &                  & PLA & Red  & R36\\
K1,L4,ESO28SC15&                 0:21:27 & -73:44:55 & C & 2.20 & 2.20 & - &                  & CMD & 9.49 & R18\\
BOLOGNA A      &                 0:21:31 & -71:56:07 & C & 0.80 & 0.80 & - & sup 47 Tucanae   & CMD & IAC  & R56\\
L5,ESO28SC16   &                 0:22:40 & -75:04:29 & C & 1.10 & 1.10 & - & Old IAC          & CMD & 9.61 & R18\\
\hline
\multicolumn{11}{c}{LMC Star Clusters}\\
\hline
NGC1466,SL1,LW1,ESO54SC16,KMHK1 & 3:44:33 & -71:40:17 & C & 3.50 & 3.50 & -  & Globular Cluster & CMD& 10.17&  R41\\
SL2,LW2,KMHK2                   & 4:24:09 & -72:34:23 & C & 1.60 & 1.60 & -  &                  & PLA&  Red&  R55\\
KMHK3                           & 4:29:34 & -68:21:22 & C & 0.80 & 0.80 & -  &                  & PLA&  Red&  R55\\
NGC1629,SL3,LW3,ESO55SC24,KMHK4 & 4:29:36 & -71:50:18 & C & 1.70 & 1.70 & -  &                  & COL&  Red&  R5\\
HS8,KMHK5                       & 4:30:39 & -66:57:25 & C & 0.80 & 0.80 & 30 &                  & PLA&  Red&  R55\\
SL4,LW4,KMHK7                   & 4:32:38 & -72:20:27 & C & 1.70 & 1.70 & -  &                  & CMD&  9.23&  R6\\
KMHK6                           & 4:32:48 & -71:27:30 & C & 0.60 & 0.55 & 80 &                  & PLA&  Red&  R55\\
\hline
\end{tabular}
\begin{list}{Table Notes.}
\item Cols.~5 and 6: semimajor axes $a$ and $b$; Col.~7: position angle; Col.~9: old age 
method. Col.~10: $\rm\log(Age)$ when available, or age class. References (Col.~11):
R3: \citet{DaCosta99}; R5: \citet{Bica96}; R6: \citet{Geisler97} - relative ages; 
R12: \citet{Crowl01}; R18: \citet{Piatti05}; R26: \citet{Glatt08}; R36: \citet{Bruck75},
\citet{Bruck76}; R41: \citet{Piatti09}; R55: present paper - red (old) cluster by plate 
inspection; R56: \citet{BPF05}.
\end{list}
\end{table*}

We employed observed (\citealt{RafZar2005}), reddening-corrected (\citealt{Hunter03}) integrated
colours, and SWB types to identify old clusters (typically SWB IVB or later, \citealt{Bica96}), 
for clusters that still lack CMD ages. We also included results from integrated spectroscopy
(\citealt{Ahumada2002}). By inspection of DSS and XDSS images we excluded clusters with apparent
contamination by relatively bright stars, concerning integrated colours and spectra. We found 
41 and 117 old clusters in the SMC and LMC respectively from integrated colours 
(Table~\ref{tab2}).

\begin{figure}
\resizebox{\hsize}{!}{\includegraphics{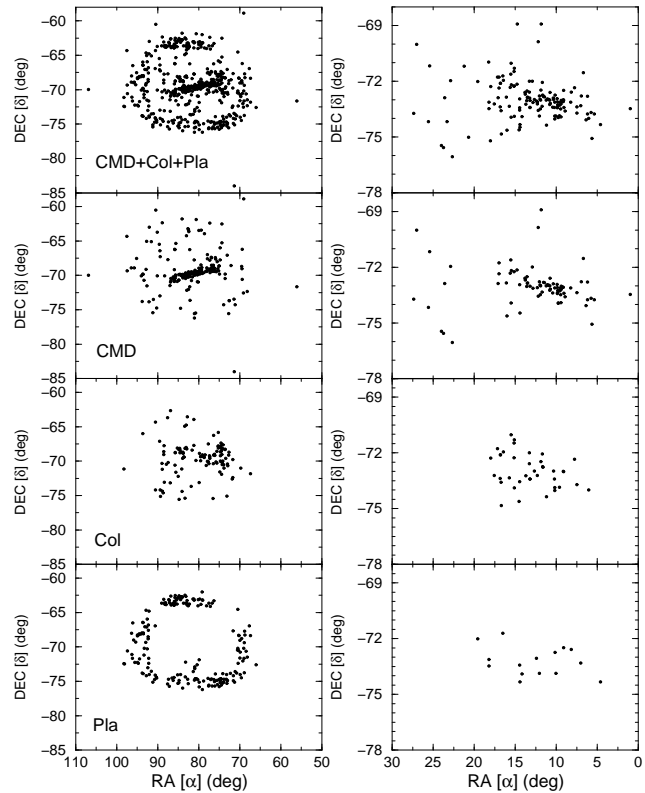}}
\caption{Angular distribution of the LMC (left panels) and SMC (right) old star 
clusters with age obtained by means of CMDs, integrated colours and plate 
inspection.}
\label{fig2}
\end{figure}

Finally, following \citet{Bruck75} and \citet{Bruck76}, we examined blue and red DSS 
and XDSS images, and ESO film sky survey plates to identify red clusters. It is remarkable 
how the red SMC clusters by \citet{Bruck76} --- his types T1 and T2 --- have been confirmed 
as old clusters by means of deep CMDs. Br\"uck disposed of U plates to help the classification.
We dispose of blue and red plates, where it was basically possible to recognise clusters with  
brighter stars red from the RGB or bluer MS stars. Also, blue clusters have as rule more 
irregular angular distributions. Most clusters that we examined by this simple method are in 
the outer parts of the LMC. By means of integrated colours, \citet{Bica96} found that the outer 
LMC disk appears to be essentially composed of old clusters. Our goal here is to provide a 
sample of probable red clusters suitable to correlation function tests, and to isolate that 
sample for CMD studies in view of VISTA and other large telescopes. There are 19 and 203 
clusters, respectively in the SMC and LMC, that are probably old (red) from our plate inspections.

Table~\ref{tab2} also includes rather populous clusters that have CMDs, integrated colours 
or plate diagnostics pointing to a blue-red  transition cluster, that occurs around 500 Myr. 
The sudden or perhaps rather smooth integrated colour change is expected from the so-called 
AGB and RGB phase transitions (e.g. \citealt{Mucc2006} and references therein). The present 
sample is a new one for such purposes. To minimise ambiguous age determinations, we have not 
included the blue-red transition clusters in our spatial correlation study, although they are, 
in principle, also old enough for dynamical purposes. Also, we note that \citet{Bica96} 
decontaminated the clusters containing atypical bright stars superimposed. We examined all clusters
showing red colours from \citet{Hunter03} and \citet{RafZar2005} on sky survey plates, and  
excluded those that appeared to be dominated by one or a few bright stars. We may have 
excluded some faint intrinsic old clusters with one or a couple bright AGB stars, but such stars 
are rare even in populous Magellanic Cloud clusters (e.g. \citealt{AM82}).

In summary, there are 522 star clusters in the LMC that can be currently considered as old 
as, or older than the Hyades. The SMC contains 145 such cases. Considering the LMC and SMC 
together, the total sample of old/red clusters corresponds to a fraction of $\approx13\%$ of 
the cluster-like structures (Table~\ref{tab2}). Details on the angular distribution of this
sub-sample are shown in Fig.~\ref{fig2}. The old CMD sample shows a well-defined LMC bar, 
while The SMC probably shows a thick edge-on disk (\citealt{UpCat} and references therein). 
Red integrated colours complement these samples mostly for fainter clusters. The LMC plate  
sample corresponds essentially to the outer disk. We emphasise that the present sharp 
inner border of the old sample is an artifact, but not the outer ring structure, as can be 
seen in the most recent census of clusters and related objects (\citealt{UpCat}). The outer
LMC disk ring is a real feature, probably produced as a consequence of the last LMC/SMC 
encounter that took place $\approx200$\,Myr ago (\citealt{BekChi07}). This structure is 
present in the uniform plate survey by \citet{Kontizas90} and in that by \citet{BSDO99}. 
The magnitude-limited integrated photometry of LMC clusters by \citet{BH95} also showed 
this structure for the oldest age group. In the present study, essentially all known red 
clusters in the outer LMC are included in that locus. The geometries of the sub-samples 
were established by each survey, but have apparently not affected the correlation functions,
as shown by the tests with different old cluster subsamples (Sect.~\ref{2pcf}).

In their surveys, \citet{BH95}, \citet{BSDO99}, and \citet{BD00} employed film copies of 
the ESO Schmidt telescope Red Survey and the UK Schmidt telescope SERC-J (blue band)  
survey in Australia\footnote{http://www.roe.ac.uk/ifa/wfau/ukstu/platelib.html\#UKSTmc}. 
The red plates trace emission nebulae by means of H$\alpha$. The limiting magnitudes are 
R=21.5 and Bj=22.5, respectively. Thus the detection limit of stars in clusters is very 
deep, especially in J. However, only the new generation of CCD surveys will permit to 
quantify the completeness of those samples compiled or discovered by our group, and 
certainly to explore an as yet undetected population of fainter objects. 

\section{Size distribution functions: structural hierarchy}
\label{DistribFunc}

The spatial distribution of interstellar gas follows a fractal structure ranging 
over many scales, from the subparsec at the smallest to the cluster and star complexes 
at the largest. This suggests that, if stars are formed mostly in the densest regions,
they should also form in fractal patterns (e.g. \citealt{ElmeElme01} and references 
therein). Indeed, the power-law nature of the size distribution function has been
observed in Galactic giant molecular clouds (\citealt{ElmeFal96}) and cloud clumps 
(\citealt{WBS95}). The end result of this process is that the cluster size distribution 
should follow a declining power-law with size, a behaviour that has been observed in 
several galaxies (e.g. \citealt{ElmeSalz99}; \citealt{ElmeKauf01}; \citealt{BGL05}). 

Our first approach in the investigation of the hierarchical structures in the Clouds 
is by means of the size (i.e. absolute radius, \rcl) distribution functions
$\phi(\rcl)=dN/d\rcl$. We build $\phi(\rcl)$ individually for all classes of objects 
listed in Table~\ref{tab1} based on the apparent major and minor axes given in the 
updated catalogue, together with the Cloud distances $d_{\rm LMC}\approx50$\,kpc and 
$d_{\rm SMC}\approx60$\,kpc (e.g. \citealt{Schaefer08}). Although the disks of
both Clouds are inclined with respect to the line of sight, the effect of the 
distance correction on the absolute radius distribution is small, to within the
error bars (see App.~\ref{apxA}).

\begin{figure}
\resizebox{\hsize}{!}{\includegraphics{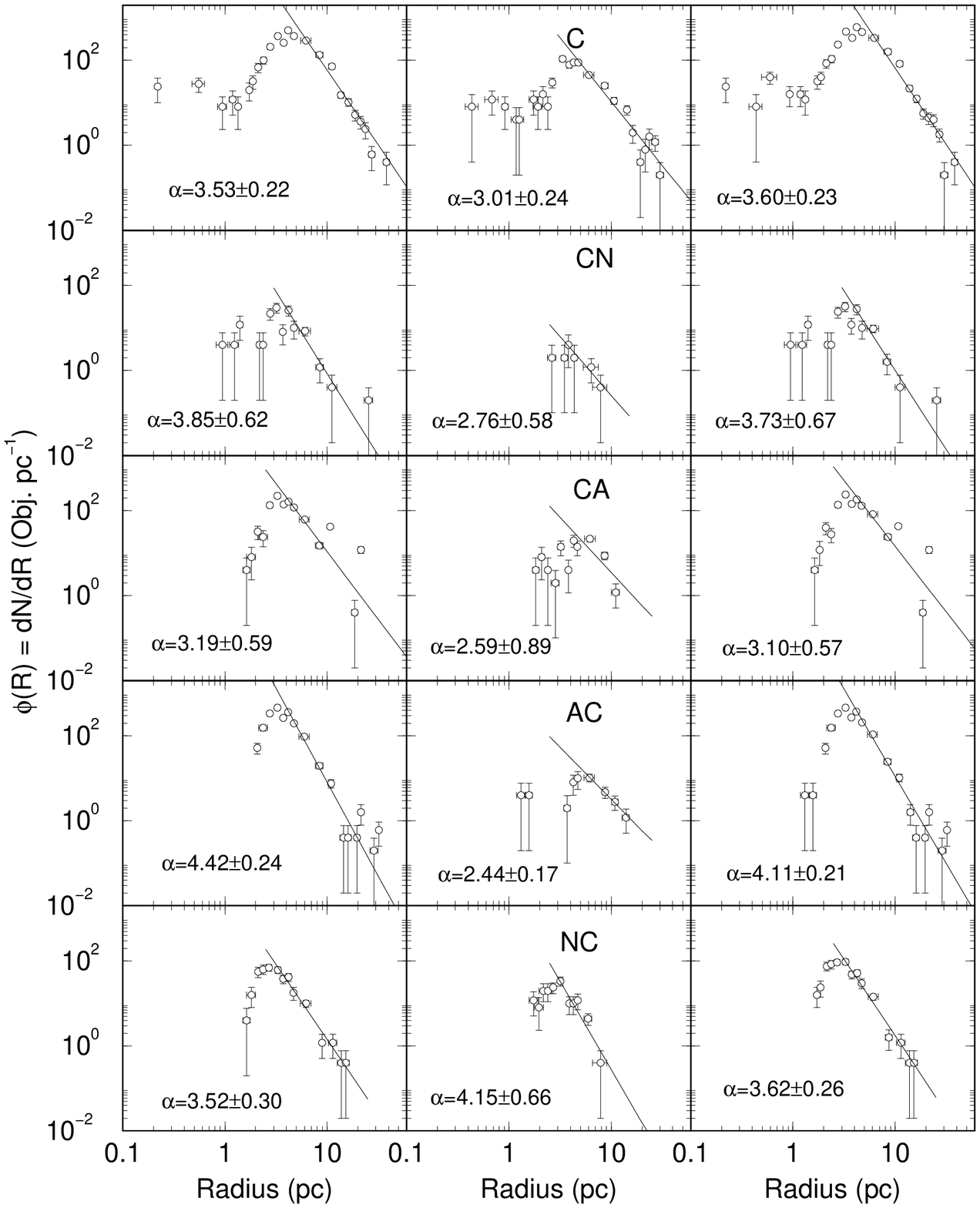}}
\caption{Radius distribution function of the cluster-like structures in the LMC
(left panels), SMC (middle), and both Clouds combined (right). The radial range
not affected by incompleteness is fitted with the power-law $\phi(R)\propto R^{-\alpha}$ 
(solid line).}
\label{fig3}
\end{figure}

\begin{figure}
\resizebox{\hsize}{!}{\includegraphics{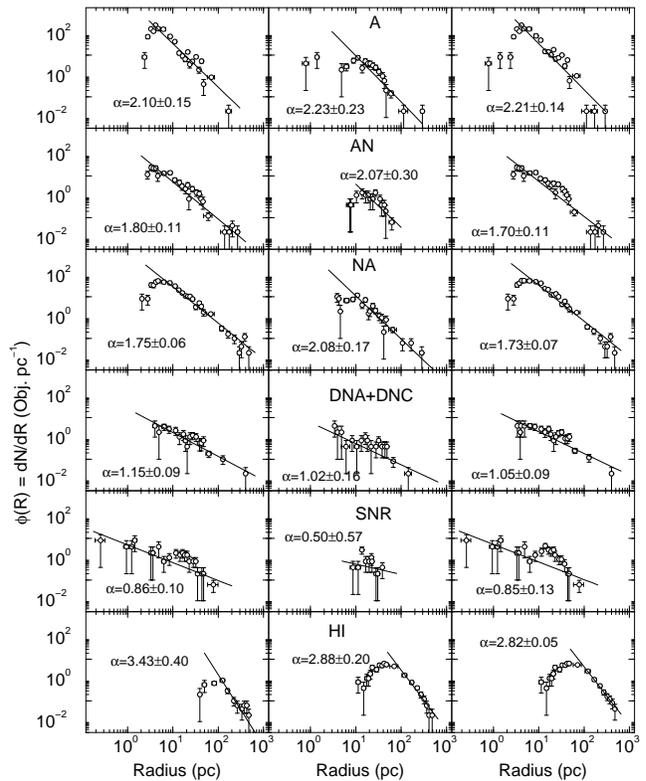}}
\caption{Same as Fig.~\ref{fig3} for the non-Cluster structures. Besides being
structures apart, SNR and HI shells have distributions significantly different from 
those of the cluster-like and non-clusters.}
\label{fig4}
\end{figure}

The radius distribution functions for the cluster-like structures (Fig.~\ref{fig3})
are characterised by a steep decline for $\rcl\ga4$\,pc, which corresponds to
about 0.25\arcmin. As we discuss in App.~\ref{apxB}, observational incompleteness 
probably accounts for the shape of $\phi(\rcl)$ in the small-size range, $\rcl\la4$\,pc.
The maximum radius ($R_{max}$) reached by the cluster-like structures in the LMC
is $R_{max}\approx40$\,pc, and $R_{max}\approx30$\,pc in the SMC. Power-law fits 
($\phi(\rcl)\propto\rcl^{-\alpha}$) to the incompleteness-unaffected range are 
obtained with rather steep slopes, $\alpha\ga3$, especially for the (statistically) 
well-defined distributions. The values of $R_{max}$ and $\alpha$ are given in 
Table~\ref{tab1}.

\begin{figure}
\resizebox{\hsize}{!}{\includegraphics{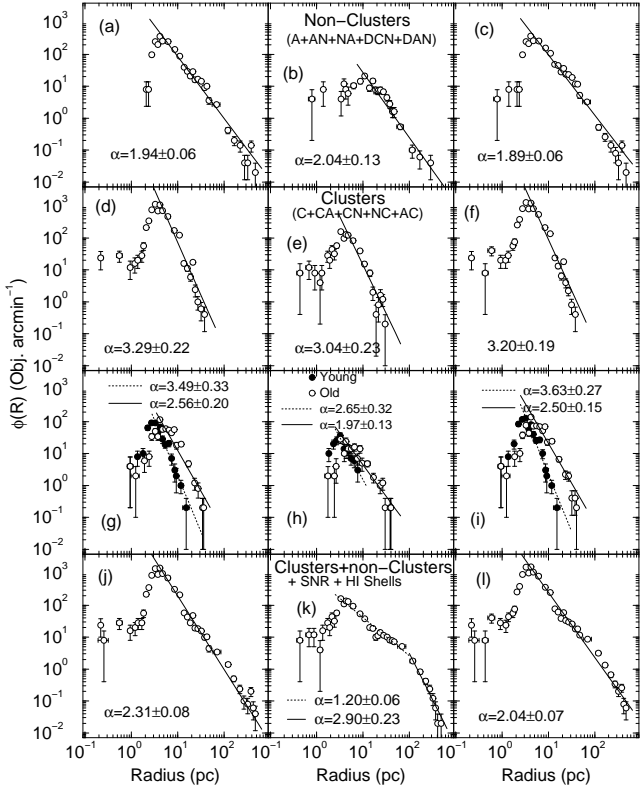}}
\caption{Same as Fig.~\ref{fig3} for the composite distribution functions of the 
non-clusters (panels a - c) and clusters (d - f). The sub-samples of the young and
old clusters are in panels (g) - (i). All structures together, including SNR and
HI shells, are shown in the bottom panels.}
\label{fig5}
\end{figure}

Figure~\ref{fig4} shows $\phi(\rcl)$ for the remaining object classes. Except for
the SNR, the distributions are qualitatively similar to those in Fig.~\ref{fig3}, with
a decline for large radii. However, compared to the cluster-like classes, the 
power-law slopes are significantly shallower ($\alpha\la2$) and the maximum radii 
are $\approx10$ times as large, reaching $R_{max}\approx500$\,pc in the LMC and 
$R_{max}\approx300$\,pc in the SMC. The incompleteness-related turnover for the A, AN,
and NA classes occurs at the same radii as that in the cluster-like objects. The
HI shells, on the other hand, have a turnover at $\rcl\approx70$\,pc 
($\approx4.4\arcmin$), which might reflect a real effect, not related to
completeness. 

Based on similarities of $R_{max}$ and $\alpha$, we define the C, CA, CN, NC, and AC 
classes as cluster-like structures, while A, AN, NA, DNC, and DAN as non-clusters.
Their composite radius distributions are shown in Fig.~\ref{fig5}, together with the 
power-law fit.
 
As expected from the above discussion, the cluster-like slopes for the LMC, SMC,
and LMC$+$SMC distributions ($\alpha\approx3$) are significantly steeper than the 
corresponding ones derived for the non-clusters ($\alpha\approx2$). Figure~\ref{fig5}
also shows the distributions obtained by adding all the structures, including the SNR
and HI shells. While most of the individual features are preserved, the SMC profile,
on the other hand, now requires two different power-laws to be described.

The slopes in the radius distribution of the non-clusters are consistent with
those measured for H\,II regions in spiral galaxies (\citealt{Oey03}). 

\subsection{Young and old clusters}
\label{YOSC}

We derive the size distribution functions of the young and old star cluster population.
We take the CN (age $\la10$\,Myr) and NC (age $\la5$\,Myr) classes (Table~\ref{tab1}) to 
represent the young star clusters. The old (age $\ga600$\,Myr) ones were selected according 
to the criteria discussed in Sect.~\ref{OSC}.

Significant differences are observed in the size distribution functions (Figure~\ref{fig5}), 
especially in the LMC. The distribution function of the young clusters falls off with radius 
at a steeper rate than the old ones, reaching a maximum size ($R_{max}\approx15$\,pc) less
than half of that reached by the old ones ($R_{max}\approx40$\,pc). In the statistically 
more significant distributions of the LMC and SMC combined (right panels), the young clusters 
fall off with the slope $\alpha\approx3.6$, while the old ones have $\alpha\approx2.5$. 
The differences in $R_{max}$ and slope probably reflect the several $10^8$\,yr of dynamical 
evolution of the old clusters, a consequence of which is an expansion of the outer parts 
(e.g. \citealt{Khalisi07}) of the clusters that survive the infant mortality (e.g. 
\citealt{GB06}) phase\footnote{However, recent evidence suggests that, during the 
infant mortality, the star cluster population has been depleted by less than $\approx30\%$,
both in the the SMC (\citealt{deGG08}) and LMC (\citealt{deGG09}).}. 

The size distribution functions of the young and old clusters fall off at a steeper rate  
($\alpha>2$) than the non-clusters ($\alpha<2$).

\subsection{A simple mass distribution function}
\label{MDF}

Hierarchically structured gas is expected to form star clusters with mass distributed 
according to a power-law of the form $dN/d\mcl\propto\mcl^{-\beta}$, with
$\beta\approx2$ (e.g. \citealt{ElmeCon}).

Below we apply a simple method to analytically transform the cluster radius 
distribution function into a mass distribution. We wish to test if our sample of 
LMC and SMC clusters basically follow the above mass distribution. We caution that 
our approach to the mass distribution is a simplification, since we do not take into 
account individual mass-to-light (M/L) ratios, which are known to vary considerably 
between young and old clusters (e.g. \citealt{BAA88}; \citealt{CB91}; \citealt{Leith99}). 
However, the presently extracted sample (\citealt{UpCat}) is by far ($\approx87\%$) 
dominated by young clusters and, thus, large variations of the M/L ratio are not 
expected. Besides, instead of computing individual masses from integrated luminosity, 
we use scaling relations that apply well to a wide variety (in terms of age, mass, and 
size) of Galactic star clusters to transform one kind of distribution function into 
another. In this sense, we expect that the adopted radius to mass transformation is 
representative of the average M/L ratio of the clusters.

We start by assuming a spherical star cluster with a mass radial density profile 
that can be described by a King-like function\footnote{Similar to the function 
introduced by \cite{King1962} to describe the surface brightness profiles in the 
central parts of globular clusters.} $\sigma_M(R)=\sigma_{M0}/(1+(R/R_c)^2)$, where 
$\sigma_{M0}$ is the surface mass density at the cluster centre and \rc\ is the core 
radius. We also consider that essentially all stars are contained within $0\leq R\leq\rcl$,
where \rcl\ is the cluster radius. The spatial mass density of such a structure can 
be computed from inversion of Abell's integral, 

$$\rho(R)=-\frac{1}{\pi}\int_R^\infty\frac{\partial\sigma_M(\chi)}{\partial\chi}
\frac{d\chi}{\sqrt{\chi^2-R^2}} = \frac{\sigma_{M0}}{2\rc}\left[\frac{1}
{1+(R/\rc)^2}\right]^{3/2}.$$

Thus, the cluster mass can be computed from

\[ \mcl\approx\int_0^{\rcl}\rho(R)4\pi\,R^2\,dR=2\pi\sigma_{M0}\rc^2\times\]
\[~~~~~~~~~~~~~~~\times\left[arcsinh(\rcl/\rc) - \frac{1}{1+(\rc/\rcl)^2}\right]. \]
 
Galactic star clusters with ages from a few Myr to $\sim1$\,Gyr, masses within 
$50\,\ms\la\mcl\la7\times10^3\,\ms$, and radii within $\rm2\,pc\la\rcl\la20$\,pc, 
have the relation between \rcl\ and \rc\ well approximated by $\rcl\approx9\rc$ (e.g. 
\citealt{LKstuff}). Comparable ratios are observed in LMC and SMC star clusters (e.g. 
\citealt{MG03a}; \citealt{MG03b}; \citealt{Carva08}). Under these assumptions we have
$\mcl\approx4\pi\sigma_{M0}\rc^2\approx0.16\,\sigma_{M0}\rcl^2$. This equation, together 
with central mass densities in the range $30\ms\,pc^{-2}\la\sigma_{M0}\la600\ms\,pc^{-2}$, 
accounts for the distribution of cluster mass and core radius (\citealt{Pi5}). Then, the 
transformation of the radius distribution to mass is given by 
$\phi(\mcl)=\phi(\rcl)/(0.31\sigma_{M0}\rcl)$, and the
average cluster mass-density is a declining function of the cluster radius, $\bar\rho(\ms\,pc^{-3})=\frac{\mcl}{(4/3)\pi\rcl^3}\approx\frac{\sigma_{M0}}{27}\rcl^{-1}$. 
For a given \rcl, the radius to mass scalings depend only on $\sigma_{M0}$ as
$\mcl\propto\sigma_{M0}$ and $\phi(\mcl)\propto\sigma_{M0}^{-1}$ that, for different 
values of $\sigma_{M0}$, preserve the shape of the mass distribution, only changing the 
mass values.

\begin{figure}
\resizebox{\hsize}{!}{\includegraphics{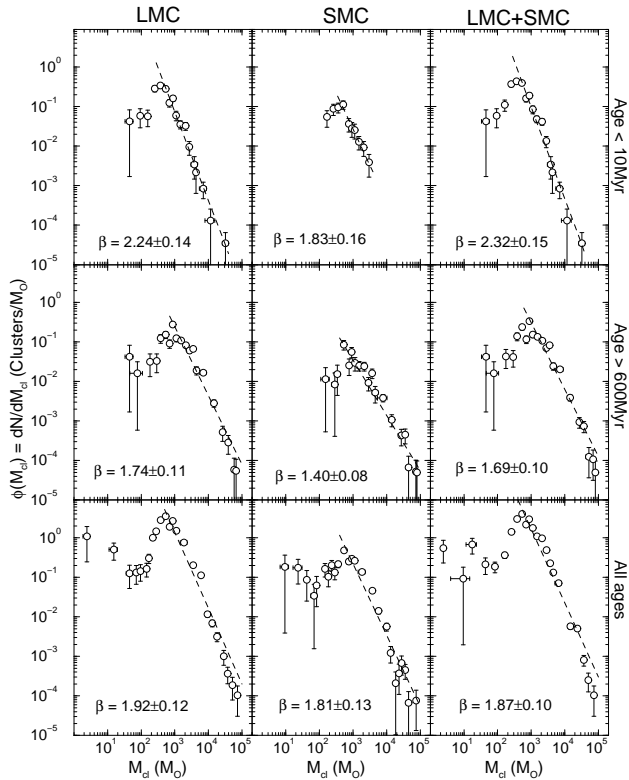}}
\caption{Estimated mass distribution functions with a power-law 
($\phi(\mcl)\propto\mcl^{-\beta}$) fitted to the large cluster mass range
of the LMC (left panels), SMC (middle), and the combined LMC$+$SMC (right)
clusters. Age ranges are $\la10$\,Myr (top panels), $\ga600$\,Myr (middle),
and all ages (bottom). }
\label{fig6}
\end{figure}

We use the above scaling relations to transform the cluster radius distribution functions 
(Fig.~\ref{fig5}) into mass distributions, $\phi(\mcl)d\mcl=\phi(\rcl)d\rcl$, separately 
for the LMC and SMC, and the combination of both, LMC$+$SMC. We also consider the age 
ranges $\la10$\,Myr, $\ga600$\,Myr, and clusters of all ages combined. The average value 
(Galactic star clusters - \citealt{Pi5}) of the central mass density, 
$\sigma_{M0}\approx300\ms\,pc^{-2}$, is used to compute the 
mass distributions (Fig.~\ref{fig6}); fit parameters are given in Table~\ref{tab3}. The 
mass distribution of the very young clusters in both Clouds falls off at a steeper rate 
towards large masses than that of the old ones, which is consistent with a mass-dependent
disruption time-scale (e.g. \citealt{Lamers05}). Also, the slopes are steeper in the LMC
than in the SMC. These slopes are consistent with those of the mass distributions of star 
clusters in different galaxies (\citealt{ElmeCon}).

\begin{table*}
\caption[]{Mass distribution properties}
\label{tab3}
\renewcommand{\tabcolsep}{1.1mm}
\renewcommand{\arraystretch}{1.25}
\begin{tabular}{crccccrccccrccc}
\hline\hline
       &\multicolumn{4}{c}{LMC}&&\multicolumn{4}{c}{SMC}&&\multicolumn{4}{c}{LMC$+$SMC}\\
       \cline{2-5}\cline{7-10}\cline{12-15}\\
Age range& N &$M_{min}$ &$M_{max}$&$\beta$&& N &$M_{min}$ & $M_{max}$&$\beta$&& N &$M_{min}$ & $M_{max}$&$\beta$\\
  (Myr)  &   & (\ms)    &   (\ms) &       &&   &  (\ms)   &   (\ms)  &       &&   &   (\ms)  &   (\ms)  &       \\
\hline
$\la10$&201 &$5.0\times10^2$&$3.2\times10^4$&$2.24\pm0.14$&&47&$5.0\times10^2$&$3.0\times10^3$
            &$1.83\pm0.16$&&198&$5.0\times10^2$&$3.2\times10^4$&$2.32\pm0.15$\\
$\ga600$&435&$9.0\times10^2$&$7.0\times10^4$&$1.74\pm0.11$&&135&$5.0\times10^2$&$8.0\times10^4$
            &$1.40\pm0.08$&&552&$9.0\times10^2$&$8.0\times10^4$&$1.69\pm0.10$\\
All ages&3700&$5.0\times10^2$&$7.0\times10^4$&$1.92\pm0.12$&&629&$5.0\times10^2$&$8.0\times10^4$
            &$1.81\pm0.13$&&4271&$5.0\times10^2$&$8.0\times10^4$&$1.87\pm0.10$\\
\hline
\end{tabular}
\begin{list}{Table Notes.}
\item N is the number of clusters used to fit the mass range $M_{min}<\mcl<M_{max}$ with
the power-law $\phi(\mcl)\propto\mcl^{-\beta}$. $M_{min}$ and $M_{max}$ computed
for $\sigma_{M0}=300\ms\,pc^{-2}$; they scale linearly with $\sigma_{M0}$.
\end{list}
\end{table*}

The mass distributions of the very young clusters are characterised by a maximum mass 
of $M_{max}\approx1.2\times10^4\,\ms$ (LMC) and $M_{max}\approx3\times10^3\,\ms$ (SMC), 
while for the old ones it is $M_{max}\approx(7-8)\times10^4\,\ms$ in both Clouds. The 
latter values are considerably higher than the maximum mass of typical Galactic open
clusters (e.g. \citealt{Piskunov07}). The decline in the number of clusters with mass 
below $M_{min}$ (Table~\ref{tab3}) is probably related to observational incompleteness 
in the detection of small clusters (Sect.~\ref{DistribFunc}). For comparison purposes 
we also show in Fig.~\ref{fig6} the mass distributions for the LMC$+$SMC clusters, as 
well as those corresponding to clusters of all ages, in which the basic features of the 
individual distributions are preserved. 

Interestingly, the maximum mass of the LMC clusters younger than $\approx30$\,Myr
(\citealt{deGA06}), and SMC ones younger than $\approx10$\,Myr (\citealt{deGG08}),
is about 2.4 times higher than that of the very young LMC and SMC clusters 
(Table~\ref{tab3}). For LMC clusters younger than $\approx5.6$\,Gyr (\citealt{deGA06}) 
and SMC ones younger than $\approx1$\,Gyr (\citealt{deGG08}), the ratio increases to 
$\approx4$. Although the somewhat different age ranges, consistency between both sets 
of $M_{max}$ values can be reached with the central mass densities 
$\sigma_{M0}\approx700\ms\,pc^{-2}$ and $\sigma_{M0}\approx1200\ms\,pc^{-2}$, 
respectively for the very young and old clusters. The somewhat higher values of
$\sigma_{M0}$ in the MCs clusters, with respect to the Galactic open clusters,
is consistent with the relative cluster mass ranges encompassed by the LMC 
(\citealt{deGA06}), SMC (\citealt{deGG08}), and Milky Way (\citealt{Piskunov07}) 
mass distributions. 

Finally, if we take into account variations of M/L with age for the 
dominant (in number) young clusters (e.g. \citealt{BC03}), the actual mass values for 
clusters younger than $\approx8$\,Myr would be $\ga30\%$ lower than the average-(M/L) 
estimates above, and $\approx20-40\%$ higher for those with age within $\approx13-30$\,Myr. 
Given that the number of clusters decreases with age (see, e.g. \citealt{deGG09}, for the
age-distribution of the MC clusters), our mass estimates in each bin of the mass 
distributions (Fig.~\ref{fig6}) may be somewhat overestimated. 

\section{Hierarchy associated with star formation}
\label{2pcf}

In a hierarchical scenario, young star clusters are expected to preserve some memory of 
the physical conditions prevailing in their birthplace. Because of random motions along
many orbits under the galactic potential, the spatial distribution of old star clusters, 
on the other hand, should be very little reminiscent of the primordial one. According to 
this scenario, the frequency of young star clusters lying relatively close to each other 
- and to star-forming structures - should be higher than for the old ones. Based on the 
590 LMC clusters ($\approx13\%$ of the present sample size - Table~\ref{tab1}) catalogued
by \citet{Bica96} with ages derived from the UBV colours by \citet{Girardi95}, 
\citet{EfreElme98} found that the average age difference between pairs of clusters
increases with the separation, which they interpreted as resulting from star formation
that is hierarchical in space and time. A similar result - and interpretation - was 
found by \citet{GalDis} for pairs of open clusters in the Milky Way disk. 

We investigate this point further by means of the degree of spatial correlation among 
groups of objects characterised by different age ranges and sizes. We use the young and 
old star clusters defined in Sect.~\ref{YOSC}. 

Consider two groups of objects, $A$ and $B$. For each object in $A$, we compute the 
angular separation with respect to all objects in $B$. After applying the same 
procedure to all objects in $A$, we build the two-point-correlation function (2PCF), 
which measures the fractional number of pairs $N$ that lie within a given separation 
$\xi$ and $\xi+d\xi$, $2PCF(\xi)\equiv dN/d\xi$. According to this definition, the 
2PCF is simply the angular separation distribution function. 

%\begin{figure}
%\resizebox{\hsize}{!}{\includegraphics{Fig7.eps}}
%\caption{Frequency distribution of the LMC (top panels) and SMC (bottom) extended
%object coordinates, with respect to the total number of objects.}
%\label{fig7}
%\end{figure}

\begin{figure}
\resizebox{\hsize}{!}{\includegraphics{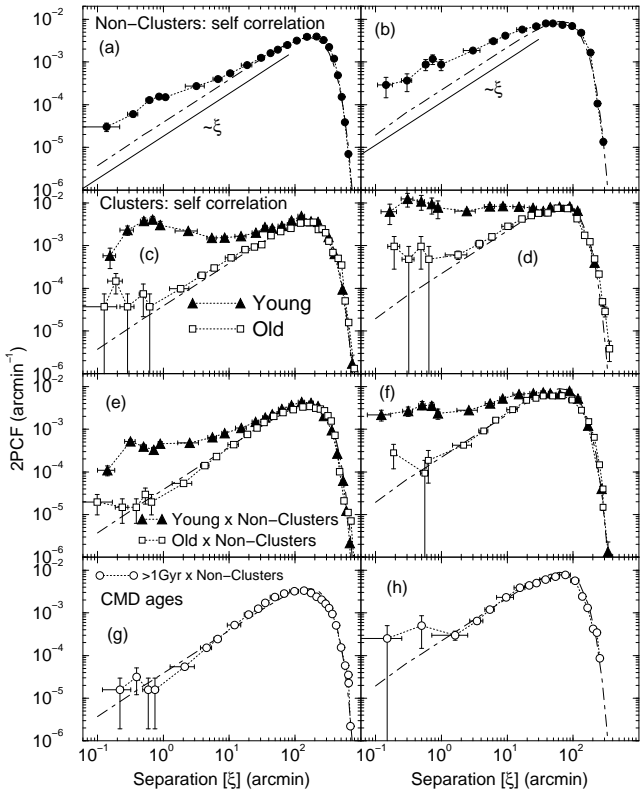}}
\caption{Two-point correlation functions for the LMC (left panels) and SMC (right)
extended objects. The simulated 2PCF (dot-dashed line) increases linearly with the
separation $\xi$. Panels (a) - (d): spatial self-correlation for the non-clusters 
and clusters. Panels (e) - (f): degree of spatial correlation of the young and old
(older than the Hyades) clusters with the non-clusters. Panels (g) - (h): same as above
for the clusters older than 1\,Gyr, with the age determined from CMDs.}
\label{fig7}
\end{figure}

\begin{figure}
\resizebox{\hsize}{!}{\includegraphics{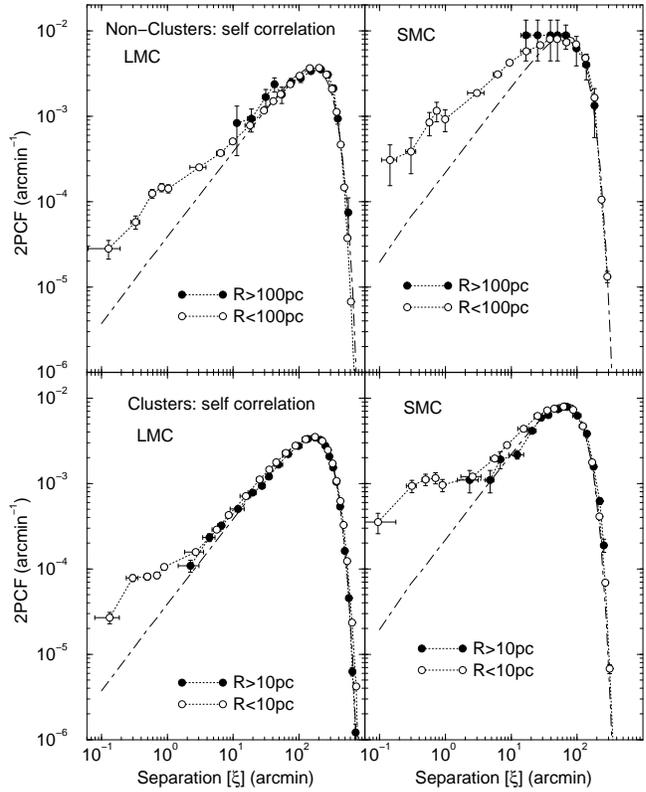}}
\caption{Same as Fig.~\ref{fig7} for the composite distribution functions of the 
non-clusters (top panels) and clusters (bottom), but differentiating for object 
size ($R$).}
\label{fig8}
\end{figure}

Artificial 2PCFs built with samples of points that emulate both the geometry 
and object distribution of the LMC and SMC are used to check the statistical 
significance of the spatial correlations. In the simulations we randomly select the 
right ascension ($\alpha$) and declination ($\delta$)coordinates  of a given point
within the actual ranges spanned by each Cloud (Fig.~\ref{fig1}) and with the same
number-frequency as the observed ones.

Irrespective of the adopted geometry, a random distribution of objects would produce 
a number of neighbours within a given separation $\xi$ that increases as 
$N(\xi)\propto\xi^2$, at least for a maximum separation $\xi_{max}$ (which should 
scale with the angular size of the simulated field). Thus, the 2PCFs should increase 
with $\xi$ as $dN/d\xi\propto \xi$. Indeed, the 2PCFs derived with the simulations 
(Fig.~\ref{fig7}) present the expected dependence with separation for 
$\xi_{max}\la80\arcmin\approx1200$ (LMC) and $\xi_{max}\la40\arcmin\approx700$\,pc 
(SMC). Beyond these values both the measured and simulated 2PCFS consistently drop, 
as a consequence of the limited size of the Clouds.

As a first step we compute the spatial self-correlation functions, in which $A=B$, 
for the LMC and SMC (Fig.~\ref{fig7}). Compared to the simulated 2PCFs, the non-clusters 
(top panels) present a relatively high degree of spatial self-correlation for separations 
smaller than $\xi\la15\arcmin\approx220$\,pc (LMC) and $\xi\la25\arcmin\approx440$\,pc 
(SMC). Young (age $\la10$\,Myr) clusters present a high degree of spatial self-correlation, 
from small to large scales (middle panels), $\xi\la35\arcmin\approx500$\,pc (LMC) and
$\xi\la25\arcmin\approx440$\,pc (SMC). Old (age $\ga600$\,Myr) clusters, on the other 
hand, have a very low degree of spatial self-correlation, restricted to separations 
$\xi\la0.6\arcmin\approx9$\,pc (LMC) and $\xi\la1.7\arcmin\approx30$\,pc (SMC). 
Interestingly, the same pattern is obtained with the 2PCFs computed for the clusters
older than 1\,Gyr, with the age determined from CMDs. Some 
degree of spatial correlation at small separations among old clusters is expected, 
since the Clouds contain binary and/or merger star clusters preferentially of comparable  
ages (e.g. \citealt{BSDO99}; \citealt{DMG02}; \citealt{Carva08}). In summary, young clusters 
have a probability of being clustered together significantly higher than old ones, both in 
the LMC and SMC. The dynamical age of clusters older than $600$\,Myr corresponds 
to $\ga4$ crossing times in the LMC, and $\ga8$ in the SMC, while for the young ones (age 
$\la10$\,Myr) it is $\la0.07$ and $\la0.13$, respectively for the LMC and SMC. Given that
a single crossing time is necessary to smear out most of the primordial structural pattern
(\citealt{GBE08}), the above conclusion is consistent with the old clusters having been 
mixed up by the random motions under the galactic potential along several $10^8$ years, 
while the young ones still trace most of the birthplace pattern. 

Now we test the degree of spatial correlation of the young and old clusters with the 
non-clusters (star-formation environments). As expected from the self-correlation analysis, 
the young clusters are highly correlated with the non-clusters (bottom panels). The
old clusters, on the other hand, appear to have some spatial correlation with the
non-clusters only at the very-small scales, $\xi\la0.2\arcmin\approx3$\,pc (LMC) and 
$\xi\la0.4\arcmin\approx7$\,pc (SMC). Part of this correlation may be due to 
projection effects on the bar. For larger separations the 2PCFs can be accounted for 
by the random distribution of objects.

\subsection{Self-correlation and object size}
\label{VwOS}

Now we examine the spatial self-correlation among clusters and non-clusters
of different radius ranges. Based on the respective radius distribution functions
(Fig.~\ref{fig5}), we take $R=10$\,pc as the boundary between small and large 
clusters; for the non-clusters we take the boundary at $R=100$\,pc. The derived 
correlation functions (Fig.~\ref{fig8}) indicate that the small clusters are more 
spatially correlated than their large counterparts. The same applies to the 
non-clusters.

Again, this picture is what should be expected from a hierarchical structure.

\subsection{Effective separation}
\label{EffSep}

Finally, we investigate the effective separation of the young and old clusters
with respect to the non-clusters. We first compute the effective separation 
($\xi_{eff}$) between a cluster and a non-cluster, which we define as the ratio 
of the angular distance ($\xi$, converted to the absolute scale) to the radius 
of the non-cluster ($R_{NC}$), $\xi_{eff}\equiv\xi/R_{NC}$. In this way, a 
cluster that is located inside a non-cluster has $\xi_{eff}<1$. 

\begin{figure}
\resizebox{\hsize}{!}{\includegraphics{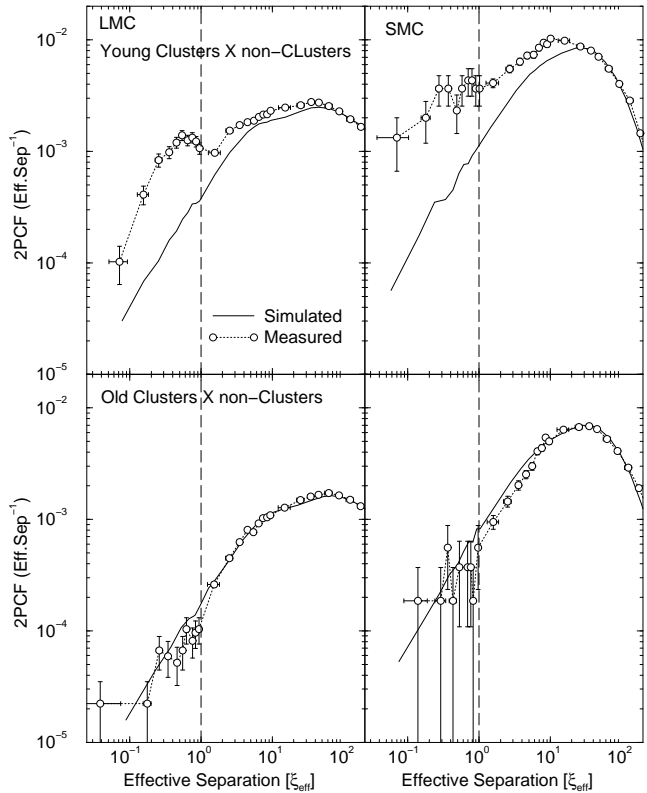}}
\caption{Two-point correlation functions of the effective separations
($\xi_{eff}\equiv\xi/R_{NC}$) between the young (top panels) and old (bottom) 
clusters with respect to the non-clusters, for the LMC (left panels) and SMC 
(right). Simulated 2PCFs are also shown (heavy-solid line). $\xi_{eff}=1$ is
indicated by the dashed line.}
\label{fig9}
\end{figure}

Two-point correlation functions built with all the $\xi_{eff}$ between samples 
$A$ and $B$ thus provide a measure of the clustering among objects in both samples. 
For comparison we use simulated 2PCFs built for object samples (young and old 
clusters and non-clusters) with coordinates selected as described in Sect.~\ref{2pcf}. 
We now also include absolute radii separately for each sample, randomly taken from 
the observed distributions (Fig.~\ref{fig5}).

The resulting 2PCFs are shown in Fig.~\ref{fig9}. Consistently with the analyses 
of the previous sections, young clusters in both Clouds present a high degree of 
clustering with the non-clusters, especially for small effective separations 
($\xi_{eff}<1$), but reaching as well high values of $\xi_{eff}$. In all scales, 
the clustering degree of the old clusters with respect to the non-clusters, on 
the other hand, can be accounted for by a random distribution of old clusters.

The above results are consistent with a strong hierarchical structuring 
of the young star clusters in both Clouds, including their time evolution 
effects. 

\section{Summary and conclusions}
\label{Conclu}

In broad lines, when star formation occurs in turbulent gas, large-scale 
structures are expected to be produced following a power-law mass distribution
($dN/dM\propto M^{-2}$ - \citealt{ElmeCon}), and with hierarchically clustered
young stellar groupings (e.g. \citealt{Efremov95}; \citealt{Elme06}).

In the present paper we address the above issue by investigating the degree of 
spatial correlation among sets of LMC and SMC extended structures, characterised 
by different properties, and its relation to star formation. Based on the catalogue 
of \citet{UpCat}, we built sub-samples that basically contain star clusters (young 
and old) and nebular complexes (and their stellar associations). The latter structures 
are related to star-forming regions; for simplicity, we refer to them as non-clusters.

In all cases (Figs.~\ref{fig3}-\ref{fig5}), the radius distribution functions follow 
a power-law ($dN/dR\propto R^{-\alpha}$) decline for large radii with slopes that 
depend on object class (and age). Taking both Clouds combined, the non-clusters 
fall-off with a slope $\alpha\approx1.9$ and reach sizes of $R_{max}\la472$\,pc. Old 
(age $\ga600$\,Myr) clusters present the somewhat steeper slope $\alpha\approx2.5$, 
while the young (age $\la10$\,Myr) ones have the steepest slope  $\alpha\approx3.6$. 
The maximum size reached by clusters is less than $\approx10\%$ of the non-clusters, 
with the old ones reaching a size $\approx3\times$ bigger than the young ones. The 
differences in slope and maximum size between the young and old clusters can be 
accounted for by long-term dynamical effects acting on the clusters. By means 
of a radius to mass scaling (Sect.~\ref{MDF}), we show that the mass distribution of
the LMC and SMC clusters follows $dN/d\mcl\propto\mcl^{-\beta}$ (Fig.~\ref{fig6}), 
with $\beta\approx2$. Within the uncertainties (Table~\ref{tab3}), this value agrees 
with the slope expected in a hierarchical scenario (\citealt{ElmeCon}). Also, the mass 
distribution for clusters younger than $\approx10$\,Myr falls off towards large masses
faster than the clusters older than $\approx600$\,Myr.  

According to the two-point correlation functions (Sect.~\ref{2pcf}), the LMC
and SMC star clusters younger than $\approx10$\,Myr present a very high degree of 
spatial correlation among themselves and, especially, with the non-clusters
(Fig.~\ref{fig7}). Clusters older than the Hyades ($\ga600$\,Myr) on the other 
hand, appear to have been heavily mixed up, probably because their ages correspond 
to several galactic crossing times and the strong perturbations associated with the 
LMC and SMC encounters (e.g. \citealt{BekChi07}). When the analysis is restricted
to clusters older than 1\,Gyr - with the age determined from CMDs, the same conclusions
are obtained. 

Considering two different radius ranges, we show that small clusters ($R<10$\,pc) 
and non-clusters ($R<100$\,pc) are spatially self-correlated, while the large ones 
are not (Fig.~\ref{fig8}). Also, young clusters in both Clouds present a very high 
degree of spatial clustering with the non-clusters, which does not occur with the 
old ones (Fig.~\ref{fig9}).

The above results, expressed in terms of the spatial and size distribution of extended 
structures in the LMC and SMC, are fully consistent with a hierarchical star-formation 
scenario, in which star complexes are part of a continuous star-formation hierarchy
that follows the gas distribution. Similar conclusions drawn from different methods 
and samples of objects have been obtained for the LMC (e.g. \citealt{ElmeEfre96};
\citealt{EfreElme98}; \citealt{HZ99}; \citealt{Livanou06}) and the SMC (e.g. 
\citealt{Livanou07}; \citealt{GBE08}).

VISTA and other large telescopes will certainly uncover a large number of faint clusters
in the Magellanic Clouds, with masses comparable to the Galactic open clusters. The same 
for embedded clusters. CMDs will provide accurate ages for them, as well as for many 
luminous, intermediate-luminosity, and low-luminosity clusters already catalogued. Also
in this context, the catalogue by \citet{UpCat} will be an essential tool, for having 
gathered and cross-identified the small and large structures in the Magellanic Clouds. 
The catalogue is also useful to help establish discoveries. The present work has provided 
as well a sub-catalogue of old and probable-old clusters, which can be useful also for 
VISTA studies.

\section*{Acknowledgements}
We thank important suggestions made by the referee, Dr. Richard de Grijs.
We thank Dr. Deidre Hunter for providing us access to the catalogue of integrated photometry
of SMC and LMC clusters. We acknowledge partial support from CNPq (Brazil). 

%--------------------------------- References -------------

\appendix
\section{Disk inclination}
\label{apxA}

The LMC and SMC disks are inclined with respect to the line of sight, and this
should introduce some variations in the absolute sizes across the disks, as
compared to the no-inclination approach adopted in Sect.~\ref{DistribFunc}. 

\begin{figure}
\resizebox{\hsize}{!}{\includegraphics{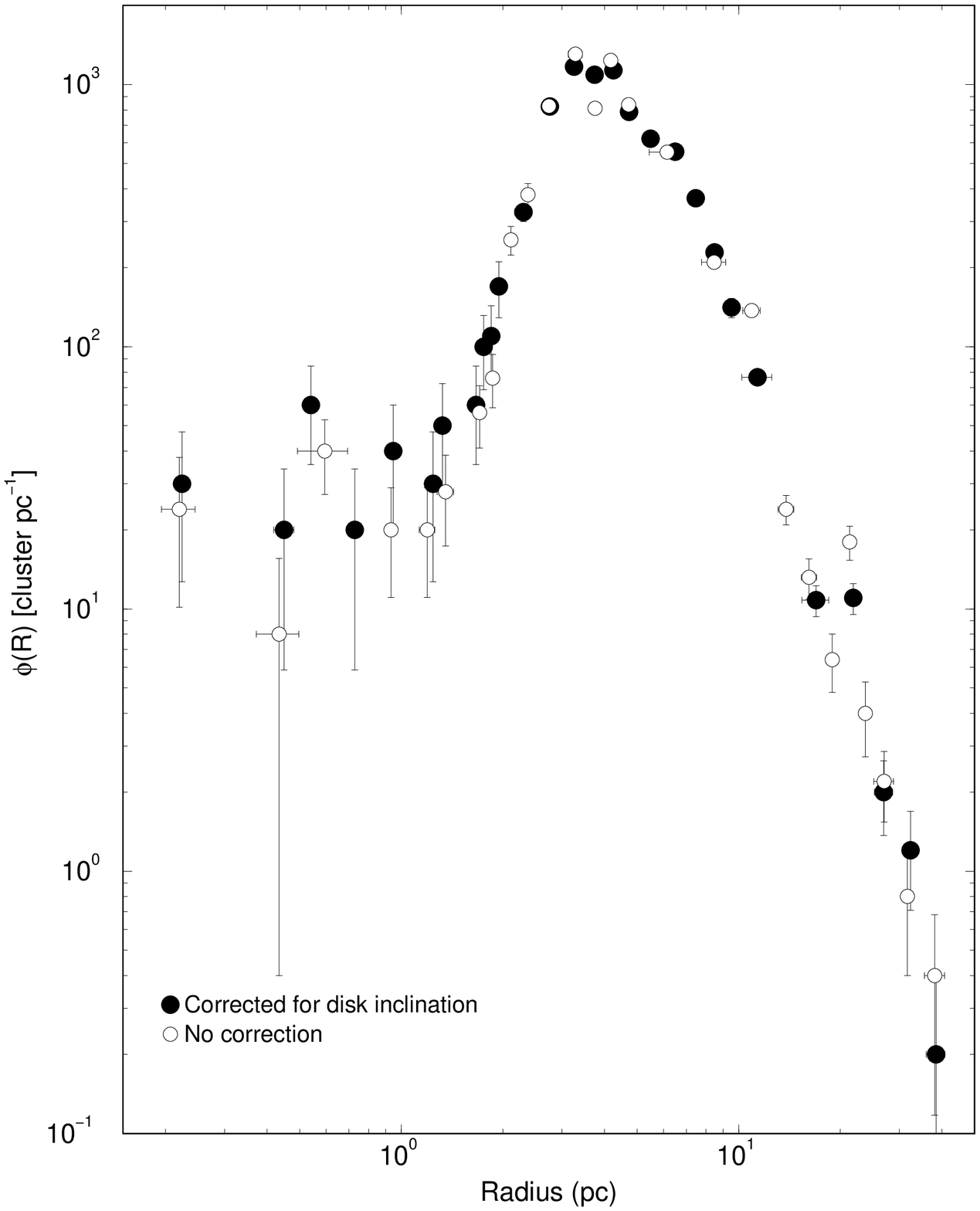}}
\caption{The inclination-corrected radius distribution function (filled symbols) 
is very similar to the uncorrected one (empty symbols).}
\label{fig11}
\end{figure}

To investigate the inclination effect on the radius distribution functions we
assume $42\degr$ (e.g. \citealt{Kontizas90}) and $40\degr$ (e.g. \citealt{SSJ04}) 
as the inclination of the LMC and SMC disks with respect to the line of sight. 
Then, the absolute size of each cluster was recomputed for its corrected distance. 
The inclination-corrected radius distribution function (for the combined LMC$+$SMC
clusters) is shown in Fig.~\ref{fig11}, in which the un-corrected distribution 
(Fig.~\ref{fig5}, panel f) is also shown for comparison purposes.
We conclude that differences are small, essentially within the error bars. This can
be accounted for by the relatively small distance corrections (with respect to the
adopted Cloud distances), and that corrections affect objects both in the near and 
far sides in the opposite sense. On average, when a large number of objects is
considered, the near and far-side corrections tend to self-compensate.

Another effect that might introduce variations on absolute cluster size is the
triaxial nature of the Clouds. The SMC, for instance, may have a line-of-sight depth 
of 6 - 12\,kpc (e.g. \citealt{Crowl01}). Thus, depth corrections would be of the 
same order as those related to inclination.

\section{Surface brightness incompleteness effects}
\label{apxB}

Since extended structures are the focus of this work, the surface brightness
(SB) incompleteness - which is expected to affect the radius distribution 
functions - should be taken into account. Basically, for a given luminosity, 
a more extended object will have on average a lower SB, and thus may not be 
detected by depth-limited surveys.

We examine this effect by means of a sample of $10^7$ artificial star clusters 
whose luminosity and radius distributions are described by $\phi(L)\,dL\propto 
L^{-2}\,dL$ and $\phi(R)\,dR\propto R^{-3.3}\,dR$, respectively. As discussed in 
\citet{ElmeElme01}, these analytical functions describe star clusters and H\,II 
regions. To reproduce the input radius and luminosity distributions, the simulated
radius and luminosity are computed from 
$R=R_m/\left[1+n_1\left((R_m/R_M)^{2.3}-1\right)\right]^{1/2.3}$ and 
$L=L_m/\left[1+n_2(L_m/L_M - 1)\right]$, where 
$R_m$, $R_M$, $L_m$, and $L_M$ are the minimum and maximum radii and luminosities, 
and $n_1$ and $n_2$ are random numbers in the range $[0.0,1.0]$.
The radius and luminosity of a given cluster are independently assigned. This process
allows that clusters of the same size (and mass) - but different ages - may 
have different luminosities, as expected from the fading lines associated with the stellar
evolution. Also, clusters with any radius within ($R_m,R_M$) are allowed to have any 
luminosity within ($L_m,L_M$). Then, the SB is computed in the usual way, 
$\mu=-2.5\log(L/\pi\,R^2)+cnt$, in arbitrary units.

\begin{figure}
\resizebox{\hsize}{!}{\includegraphics{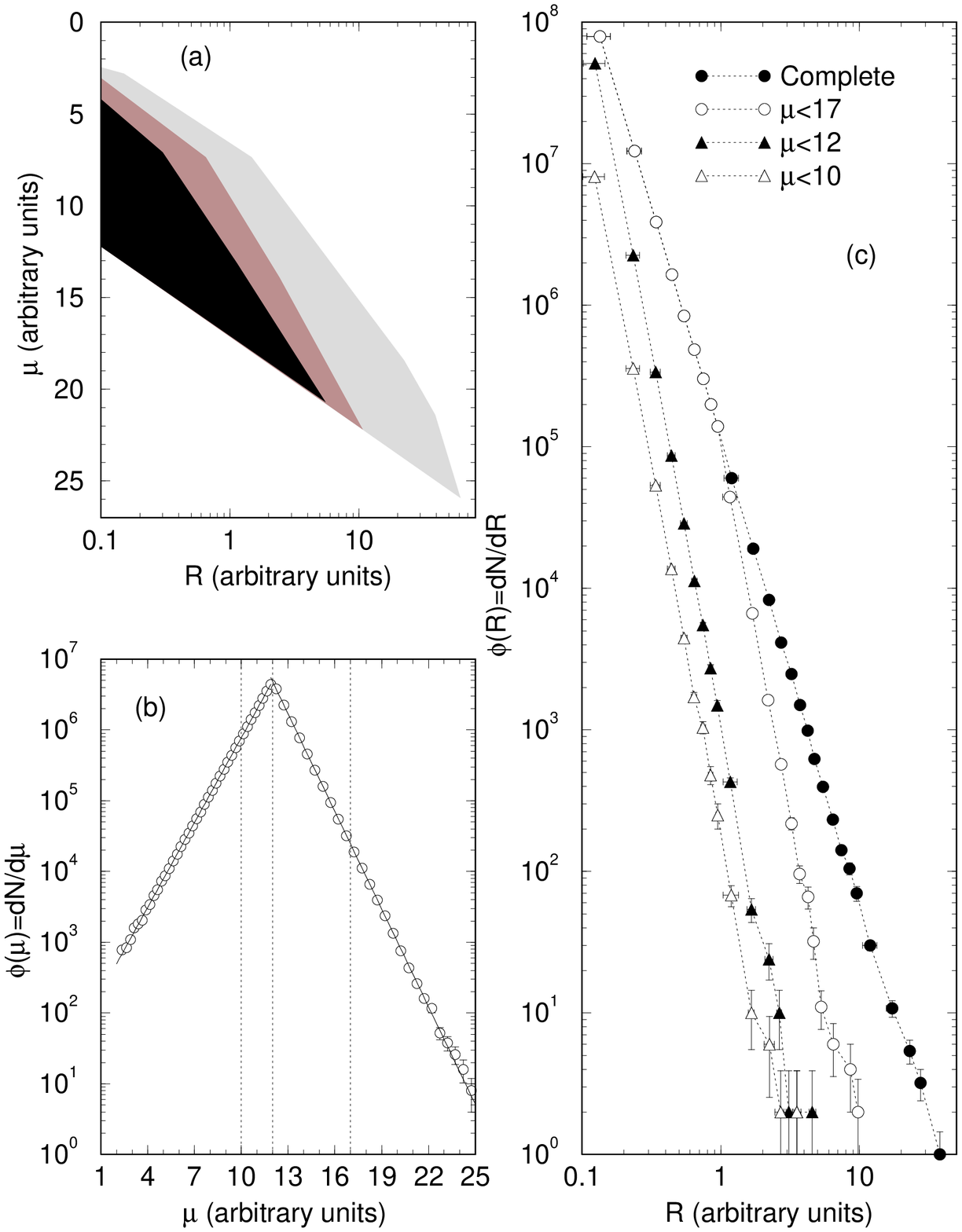}}
\caption{Panel (a): model SB distribution with respect to cluster radius; the density
of points (number of simulated clusters) roughly increases towards heavier shades of 
gray. (b): SB distribution function (fitted with exponentials - solid line) showing 
the arbitrary thresholds (dashed lines). (c): radius distributions corresponding to 
the SB cuts in (b). }
\label{fig12}
\end{figure}

The results are summarised in Fig.~\ref{fig12}. In general, the SB distribution among
the simulated clusters agrees with the expected relation of decreasing SB with cluster 
radius (panel a). The two power-laws that describe the radius and luminosity 
distributions are reflected on the shape of the SB distribution (panel b), which
first (beginning at the smallest and most luminous clusters) increases exponentially 
towards lower SBs, reaches a maximum, and falls off exponentially towards the largest 
and less luminous clusters. Based on this distribution we arbitrarily apply cuts for 
clusters with $\mu<17,~12,$ and $10$ and compute the corresponding radius distributions
(panel c). Clearly, the SB cuts preserve the power-law character of the radius
distribution. The major effect is a steepening of the slope. Indeed,while the complete
radius distribution is a power-law of slope $\alpha=-3.3$, the SB-restricted
distributions have $\alpha=-3.6\pm0.1,~-5.0\pm0.1$, and $\alpha=-5.2\pm0.1$, 
respectively for $\mu<17,~12,$ and $10$.

In summary, SB-related incompleteness affects the radius distributions preferentially
at the large-clusters tail, having little effect on the small clusters. Also, it 
preserves the power-law character of the radius distribution and, due to the 
preferential effect on large clusters, it produces a steepening of the distributions. 
Thus, the decrease in the observed radius distributions towards small clusters 
(Figs.~\ref{fig3} - \ref{fig5}) cannot be accounted for by SB incompleteness, and 
appears to be linked to an observational effect. In this sense, VISTA will be important 
also to explore the structure of small clusters in the Clouds, and to investigate
the shape of the radius distribution at the small scales. 

\end{document}